\documentclass{iau} 
\usepackage{graphicx}

\title[The missing mass conundrum of post-CE PNe] %% give here short title %%
{The missing mass conundrum of \\ post-common-envelope \\ planetary nebulae}

\author[Miguel Santander-Garc\'ia \etal]   %% give here short author list %%
{Miguel Santander-Garc\'ia$^1$,
David Jones$^{2,3}$,
Javier Alcolea$^1$,
Roger Wesson$^{4,5}$
 \and Valent\'in Bujarrabal$^1$}

\affiliation{$^1$Observatorio Astron\'omico Nacional,
Alfonso XII, 3, 28014, Madrid, Spain \\ email: {\tt m.santander@oan.es} \\[\affilskip]
$^2$Instituto de Astrof\'i­sica de Canarias, E-38205 La Laguna, Tenerife, Spain \\[\affilskip]
$^3$Departamento de Astrof\'i­sica, Universidad de La Laguna, E-38206 La Laguna, Tenerife, Spain \\[\affilskip]
$^4$Department of Physics and Astronomy, University College London, Gower St, London, UK \\[\affilskip]
$^5$European Southern Observatory, Alonso de C\'ordova 3107, Casilla 19001, Santiago, Chile
}

\pubyear{2018}
\volume{xxx}  %% insert here IAU Symposium No.
\jname{Why Galaxies Care About AGB Stars. A Continuing Challenge through Cosmic Time}
\editors{XXX, eds.}
\begin{document}

\newcommand{\msun}{\mbox{$M_{\odot}$}}

\maketitle

\begin{abstract}
Most planetary nebulae (PNe) show beautiful, axisymmetric morphologies despite their progenitor stars being essentially spherical. Angular momentum provided by a close binary companion is widely invoked as the main agent that would help eject an axisymmetric nebula, after a brief phase of engulfment of the secondary within the envelope of the Asymptotic Giant Branch (AGB) star, known as a common envelope (CE). The evolution of the AGB would be thus interrupted abruptly, its (still quite) massive envelope fully ejected to form the PN, which should be more massive than a PN coming from the same star were it single. We test this hypothesis by deriving the ionised+molecular masses of a pilot sample of post-CE PNe and comparing them to a regular PNe sample. We find the mass of post-CE PNe to be actually lower, on average, than their regular counterparts, raising some doubts on our understanding of these intriguing objects.
\end{abstract}

\firstsection % if your document starts with a section,

\section{Introduction}

Most planetary nebulae (PNe) show beautiful, aspherical morphologies with high degrees of symmetry, despite their progenitor stars being essentially spherical. The mechanism behind their shaping, however, is still poorly understood (e.g. \cite{balick02}). Angular momentum provided by a close binary companion has been widely invoked as the main shaping agent that would eject an axisymmetric nebula (\cite{jones17}). 

The mechanism in close binary systems is thought to be as follows: a star undergoing the Asymptotic Giant Branch (AGB) stage engulfs a companion via Roche-lobe overflow as it expands during the Asymptotic Giant Branch (AGB) phase. The system then undergoes a very brief ($\sim$1 year) common-envelope (CE) stage, where the evolution of the AGB star is abruptly interrupted. Spiraling-in of the secondary and drag forces would then lead to the ejection and shaping of this CE into a bipolar PN whose equator would be coincident with the orbital plane of the binary star, as happens to occur in every single case analysed so far (\cite{hillwig16}).

%This hypothesis has been progressively gaining ground, particularly with the recent discovery of a few dozens of PNe in binary systems with orbital separations orders of magnitude smaller than the typical radius of an AGB star (e.g. \cite{boffin11}, \cite{boffin12}, \cite{jones14}, \cite{corradi11}, \cite{santander15}). In fact, statistical analyses show that (at least) one in five PNe derives from a post-CE binary system (\cite{han95}, \cite{miszalski09})

On theoretical grounds, however, the physics of the CE ``friction'' and ejection processes remain very elusive. Simulations show most of the gas to be ejected along the equatorial plane, but are unable to gravitationally unbind the whole envelope of the AGB (e.g. \cite{huarte12},\cite{garciasegura18}). An exception would imply tapping energy from atomic recombination in the envelope (e.g. \cite{ohlmann16}), but then the achieved expansion velocities would likely be too large.

This draws a somewhat uncomfortable big picture: we simply do not understand the physics lying behind the death of a significant fraction of stars in the Universe.

\bigskip

{\underline{\it Single star vs. CE evolution: the total nebular mass}}. It can be argued that CE evolution implies significant differences in the mass-loss history of the primary star. 

Let us consider a single AGB star on its way to produce a PN. Most of its envelope's mass is slowly lost along the AGB evolution, and gets too diluted in the Interstellar Medium (ISM) to be detected. In contrast, the mass lost by the star during the superwind phase (last $\sim$500-3000 years), which amounts to $\sim$0.1-0.6 \msun\ for a 1.5 \msun\ star (see review by \cite{hofner18}), will form the nebula visible during the PN stage.

On the other hand, let us consider the same AGB star, but now as part of a binary system close enough to engulf its companion and undergo a CE stage. AGB engulfment will thus occur during the last few ($\sim$1-20) million years of the AGB stage (e.g. Fig. 1 in \cite{macleod12}), effectively interrupting the evolution of the star. All the mass the star did not lose into the ISM during these last million years will be present in the CE, and therefore will {\it also} be part of the PN as it is suddenly ejected.

In other words, despite the large uncertainties in the mass-loss history along the AGB, {\it PNe arising from CE events should, on average, be more massive than their single star counterparts}.

This additional mass should be detectable, as it will be close to the central stars during the lifetime of the PN, as opposed to the single star case, where it will be long gone, diluted into the ISM. Testing this hypothesis would lead to a better understanding of the ejection process. Nevertheless, complete mass determinations of post-CE PNe are virtually nonexistent. We hereby present the results of a pilot survey of this kind.

\section{Sample and Observations}

Our pilot sample is composed of 10 post-CE PNe, which amount roughly to 1/6$^\mathrm{th}$ of the total currently known. It covers a broad range of kinematical ages, central star effective temperatures and luminosities, orbital periods and morphologies. These objects are PM~1-23, Abell 41, Hen~2-428, ETHOS~1, NGC~6778, Abell~63, the Necklace, V~458 Vul, Ou~5, and NGC 2346. They lacked any attempt at detecting their molecular content by means of radioastronomical observations, except for NGC~6778 (undetected by \cite{huggins89}), and NGC 2346, already known to host a massive molecular envelope (e.g. \cite{bachiller89}). We therefore carried out spectral observations of the sample (except NGC~2346), in search for $^{12}$CO and $^{13}$CO $J$=1-0 and $J$=2-1 emission, using EMIR in the IRAM 30m radiotelescope. The angular size of the objects of the sample is generally well suited to the telescope Half Power Beam Width at the observed frequencies.

We complemented the mm-range data with archival H$\alpha$ images and optical spectra of the sample, from various telescopes and instruments, to derive their ionised masses.

%The spectral profiles of the CO emission will provide us with morphological information such as hints of equatorial or polar ejections, keplerian or expanding discs or jets, which will further constrain models of formation (e.g. \cite{soker17}). In the case that the emission is bright enough to detect other molecules (such as transitions from CO+, CN and SiO, which are included in the observed spectral ranges), we will study the chemistry of post-CE evolution in the different objects. 

%By means of similar proposals in other spectral regimes (e.g. optical) we plan to gather as much information as possible about the masses of the different components of these planetary nebulae, with the goal of testing the hypothesis that post-CE PNe should be more massive than their counterparts (for a given primary star initial mass).

%Finally, even non-detections will serve a useful purpose, as we will be able to provide upper mass limits (of the order of a few thousandths of a solar mass) for the molecular content of these PNe, which will help improve models. Also, finding a number of post-CE PNe with masses (ionised+neutral+molecular) much lower than the mass of the AGB envelope would pose a serious challenge for the current paradigm, as it would imply that most of the mass supposedly ejected from the AGB is unexpectedly missing.

\section{Results}

{\underline{\it Molecular content}}. No object was detected in $^{12}$CO or $^{13}$CO down to a {\it rms} sensitivity limit in the range 6-25 mK at 230 GHz, except for NGC~6778. This PN shows a simple, broad  $^{12}$CO $J$=1-0 emission profile, as well as double-peaked emission profiles in $^{12}$CO and $^{13}$CO $J$=2-1, whose kinematics correspond to the broken, equatorial ring investigated by \cite{guerrero12}. The peak intensity relations lead us to conclude that the $^{12}$CO $J$=1-0 is optically thin, and the excitation temperatures relatively low. Further analysis of these profiles and the excitation conditions in this nebula will be presented in Santander-Garc\'ia (in preparation).

The $^{12}$CO $J$=1-0 profile of NGC~6778 allows us to derive a molecular mass of 5$\times$10$^{-4}$ \msun\ (at 1~kpc) for this PNe by assuming a representative value of the $^{12}$CO abundance of 3$\times$10$^{-4}$. On the other hand, the sensitivities achieved in the rest of the observations allow us to derive conservative (3-$\sigma$) upper limits for the molecular masses of the other objects in the sample.

{\underline{\it Ionised content}}. The ionised mass of NGC~6778, NGC~2346, Abell~41, ETHOS~1, Hen~2-428 and PM~1-23 were derived from their H$\beta$ fluxes and apparent sizes extracted from archival data. Assumptions about the electronic temperatures were made where necessary, in order to produce conservative estimates of the ionised masses of these nebulae (i.e. largest T$_\mathrm{e}$ wherever more than one was available). Ionised masses of the Necklace, Abell~63, Ou~5, and V458~Vul were obtained from \cite{corradi11}, \cite{corradi15}, \cite{corradi15}, and \cite{wesson08}, respectively.

{\underline{\it Total mass comparison at 1~kpc}}. Masses found in this work scale with the distance to the nebulae squared. Distances to PNe, however, are still poorly known. Hence, in order to do a proper comparison with PNe not undergoing CE, we must first remove this large dependance by examining the mass every PNe would have at the same distance. Figure 1 shows the ionised and molecular masses of our sample of post-CE PNe at 1kpc, together with the ionised and molecular masses of a large sample of 44 PNe selected by \cite{huggins96} in an attempt to approach a volume-limited sample, and another sample of 27 PNe in the galactic disk, whose ionised/molecular masses were determined by \cite{boffi94} and \cite{huggins89}, respectively.

Strikingly, except for NGC~2346, the total masses of the post-CE sample seem similar, if not lower, than those of regular PNe. The median mass at 1~kpc of the combined comparison samples is 0.021 \msun, whereas for the post-CE sample it is $\leq$0.0081 \msun.

\begin{figure}[b]
% \vspace*{-2.0 cm}
\begin{center}
 \includegraphics[width=10.5cm]{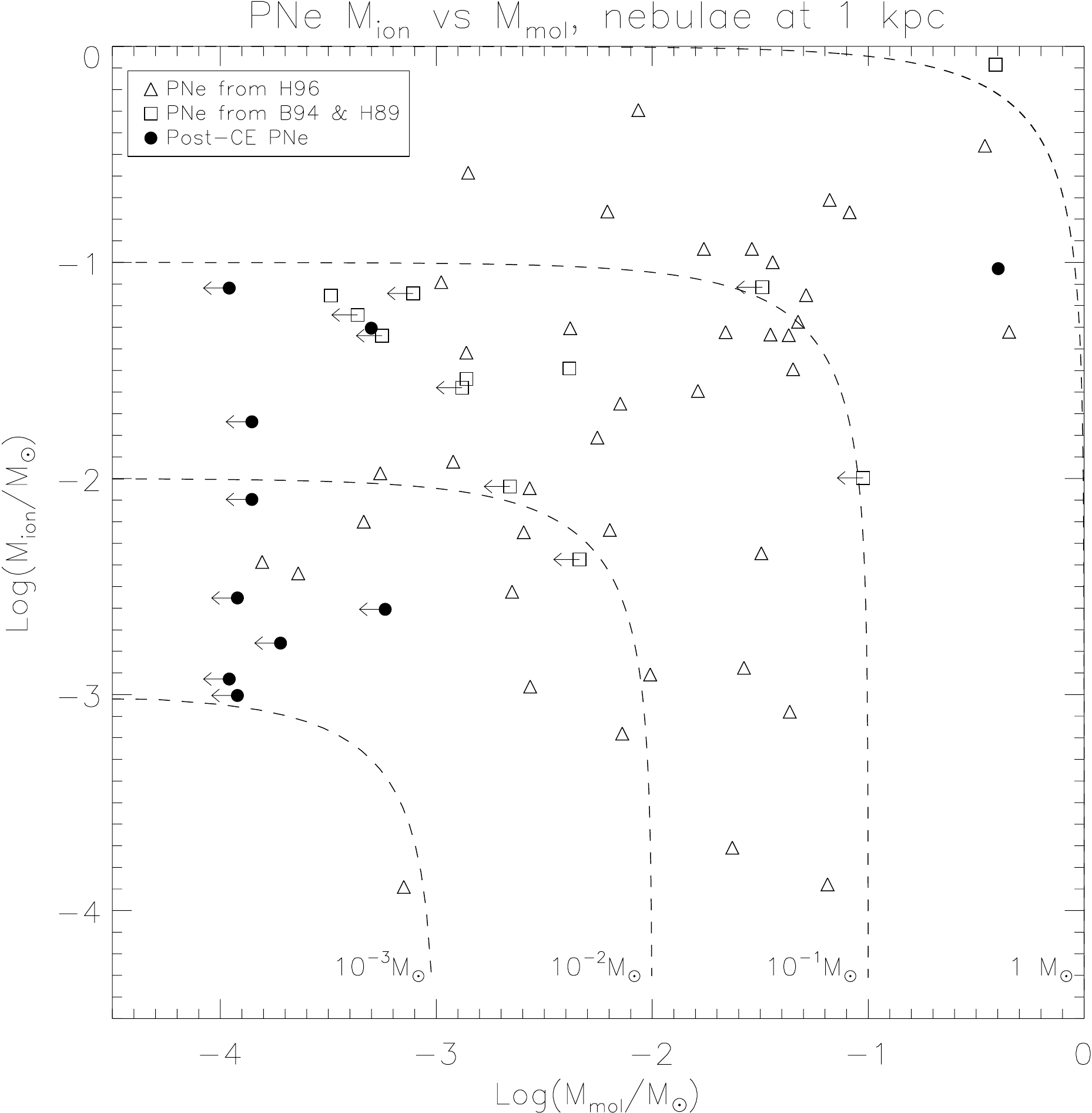} 
% \vspace*{-1.0 cm}
 \caption{Logarithmic ionised mass vs. logarithmic molecular mass at 1~kpc of our post-CE PNe sample (filled circles), PNe from Huggins \etal\ (1996) (triangles), and a combined sample from Boffi \etal\ (1994) and Huggins \etal\ (1989) (squares). Dashed lines indicate equal total (ionised+molecular) mass; individual nebulae run along these lines as their gas content is progressively ionised.}
   \label{fig1}
\end{center}
\end{figure}

\section{Conclusions}

This preliminary work provides a first indication that, contrary to expectations, post-CE PNe seem to be slightly less massive, on average, than their single star counterparts. This discrepancy could however be removed if the molecular gas of these nebulae were too cold (or hot) to be detected, or the ionised gas too hot to emit H$\alpha$, but these possibilities seem rather unlikely. Some of the mass could also be in atomic, neutral form, which has not been investigated in this work, and will be part of a future study.

On the other hand, should these results be confirmed by further observations and careful analysis of the possible biases involved, they would present us with the following interesting (and so far speculative) implications. The problem of models unable to unbind such a large mass would be less severe. A fraction of the mass could fall back forming a circumbinary disk (as in \cite{reichardt18}). If any of this material reaches the central stars, it could then be reprocessed perhaps offering an explanation for the correlation between large abundance discrepancy factors and post-CE central stars in PNe (\cite{wesson18}). We can thus wonder whether the CE itself could be not a unique, only-once process, but an episodic, recurrent one. Grazing Envelope Evolution proposed by \cite{soker15} and \cite{shiber17} could help explain such a phenomenon.


\begin{thebibliography}{}

\bibitem[Bachiller \etal\ 1989]{bachiller89}
{Bachiller, R., Planesas, P., Mart\'in-Pintado, J., \etal} 1989, \textit{A\&A}, 210, 366

\bibitem[Balick \& Frank 2002]{balick02}
{Balick, B. \& Frank, A.} 2002, \textit{ARA\&A}, 40, 439

\bibitem[Boffi \& Stanghellini (1994)]{boffi94}
{Boffi, F. R., \& Stanghellini, L.} 1994, \textit{A\&A}, 284, 248

\bibitem[Corradi \etal\ (2011)]{corradi11}
{Corradi, R. L. M., Sabin, L., Miszalski, B., \etal} 2011, \textit{MNRAS}, 410, 1349

\bibitem[Corradi \etal\ (2015)]{corradi15}
{Corradi, R. L. M., Garc\'ia-Rojas, J., Jones, D., Rodr\'iguez-Gil, P.} 2015, \textit{ApJ}, 803, 99

\bibitem[Garc\'ia-Segura, Ricker \& Taam 2018]{garciasegura18}
{Garc\'ia-Segura, G.; Ricker, P. M.; Taam, R. E.} 2018, \textit{ApJ}, 860, 19

\bibitem[Guerrero \& Miranda (2012)]{guerrero12}
{Guerrero, M. A., \& Miranda, L. F.} 2012, \textit{A\&A}, 539, 47

\bibitem[Huarte-Espinosa \etal\ 2012]{huarte12}
{Huarte-Espinosa, M., Frank, A., Balick, B., \etal} 2012, \textit{MNRAS}, 424, 2055

\bibitem[Huggins \& Healy 1989]{huggins89}
{Huggins, P. J., \& Healy, A. P.} 1989, \textit{ApJ}, 346, 201

\bibitem[Huggins \etal\ (1996)]{huggins96}
{Huggins, P. J., Bachiller, R., Cox, P., \etal} 1996, \textit{A\&A}, 315, 284

\bibitem[Hillwig \etal\ 2016]{hillwig16}
{Hillwig, T. C., Jones, D., de Marco, O., \etal} 2016, ApJ, 832, 125

\bibitem[H\"ofner \& Olofsson]{hofner18}
{H\"ofner, S.; Olofsson, H.} 2018, \textit{A\&AR}, 26, 1

\bibitem[Jones \& Boffin 2017]{jones17}
{Jones, D., \& Boffin, H. M. J.} 2017, \textit{Nature Astronomy}, 1, 117

\bibitem[MacLeod, Guillochon \& Ramirez-Ruiz 2012]{macleod12}
{MacLeod, M., Guillochon, J., Ramirez-Ruiz, E.} 2012, \textit{ApJ}, 757, 134

\bibitem[Ohlmann \etal\ 2016]{ohlmann16}
{Ohlmann, S. T.; R\"opke, F. K.; Pakmor, R.; Springel, V.; M\"uller, E.} 2016, \textit{MNRAS}, 462, L121

\bibitem[Reichardt \etal\ 2018]{reichardt18}
{Reichardt, T. A., De Marco, O., Iaconi, R.} 2018, MNRAS, submitted (\textit{arXiv}:1809.02297)

\bibitem[Soker (2015)]{soker15}
{Soker, N.} 2015, \textit{ApJ}, 800, 114

\bibitem[Shiber, Kashi, \& Soker (2017)]{shiber17}
{Shiber, Sagiv; Kashi, Amit; Soker, Noam} 2017, \textit{MNRAS}, 465, L54

\bibitem[Wesson \etal\ (2008)]{wesson08}
{Wesson, R., Barlow, M. J., Corradi, R. L. M., \etal} 2008, \textit{ApJ}, 688, L21

\bibitem[Wesson \etal\ 2018]{wesson18}
{Wesson, R., Jones, D., Garc\'ia-Rojas, J., \etal} 2018, \textit{MNRAS}, 480, 4589

\end{thebibliography}
\end{document}